\documentstyle[letter,wrapft]{ptptex}

\newcommand{\beq}{\begin{equation}}
\newcommand{\eeq}{\end{equation}}
\newcommand{\beqa}{\begin{eqnarray}}
\newcommand{\eeqa}{\end{eqnarray}}
\newcommand{\kB}{\mbox{$k_{\rm B}$}}
\newcommand{\kBT}{\mbox{$k_{\rm B}T$}}

\title{
Stiffening Transition in Vicinal Surfaces with Adsorption
}

\author{
Noriko {\sc Akutsu},\footnote{E-mail address: nori@phys.osakac.ac.jp} 
Yasuhiro {\sc Akutsu}$^{*,}$
\footnote{E-mail address: acts@phys.sci.osaka-u.ac.jp}
and Takao {\sc Yamamoto}$^{**,}$
\footnote{E-mail address: yamamoto@phys.eg.gunma-u.ac.jp}
}

\inst{
Faculty of Engineering, Osaka Electro-Communication
 University,\\
Hatsu-cho, Neyagawa, Osaka 572-8530, Japan
\\
$^*$Department of Physics, Graduate School of Science, Osaka University,\\
Machikaneyama-cho, Toyonaka, Osaka 560-0011, Japan
\\
$^{**}$Department of Physics, Faculty of Engineering, Gunma University,\\
Kiryu, Gunma 376-0052, Japan
}

\recdate{
}

\abst{
We study the vicinal surface with adsorption below the roughening temperature, using the restricted solid-on-solid model coupled with the Ising model.  We calculate the step tension $\gamma$ and the step-interaction coefficient $B$ by employing a variant of the density matrix algorithm.  We find a ``stiffening transition'' at a temperature $T_{\rm s}$ where $B$ vanishes.  At $T_{\rm s}$ surface free energy has the form $f(p)- f(0)= \gamma p + (\mbox{const.})p^5 + \cdots$ ($p$: surface gradient), which differs from the well-known $p$-$p^3$ form.
  
}

\begin{document}

\maketitle

Below the roughening temperature $T_{\rm R}$, the vicinal surface which is a slightly tilted surface relative to one of the facet plane of a crystal, is well described in terms of terraces, steps and kinks (TSK picture).  Since systems whose ``elementary excitation'' is an extended linear object (like the step), belong to the Gruber-Mullins-Pokrovsky-Talapov (GMPT) universality class,\cite{gmpt,haldane,izuyama,jayaprakash,schults,nolden,pimpinelli} we have the following form of the vicinal surface free energy (per projected area):
\begin{equation}
f(p)=f(0)+\gamma p +B p^3 +O(p^4), \label{GMPTform}
\end{equation}
where $p$ ($\propto$ step density $\rho$) is the surface gradient. This $p$-$p^3$ form of expansion is characteristic of the GMPT universality class.  Physically, $\gamma$ is the step tension, and $B$ represents the effect of step-step interactions.  To be precise, $\gamma$ and $B$ depends on the mean running direction angle (relative to one of the crystal axes on the facet plane) of the steps, which we  denote by $\theta$; we should then write $\gamma=\gamma(\theta,T)$ and $B=B(\theta,T)$ ($T$: temperature).  If we set up the $xy$ coordinates on the facet plane so that the $y$-axis corresponds to $\theta=0$ (i.e., steps are running along the $y$-direction), the angle $\theta$ relates to the surface gradients $p_{x}$ (along the $x$-direction) and $p_{y}$ (along the $y$-direction) as
\begin{equation}
p_{x}=-a_{h}\rho \cos\theta,\quad p_{y}=-a_{h}\rho\sin\theta, \label{pxy}
\end{equation}
where $a_{h}$ is the height of a single step (for convenience, we adopt the unit where $a_{h}=1$ in the following).  With (\ref{pxy}), we can regard the free energy $f$ as a function of the gradient vector ${\mib p}=(p_{x},p_{y})$, allowing us to write $f=f({\mib p})$.  For systems where the step-step interaction is short-ranged, there exists the following universal relation between $\gamma(\theta,T)$ and $B(\theta,T)$\cite{aay88,yamamoto88} (which leads to the universal Gaussian curvature jump at the facet edge\cite{aay88,yamamoto88,saam,UGCJ}):
\begin{equation}
B(\theta,T)=\frac{\pi^2 }{6} \frac{(\kBT)^2}{\tilde{\gamma}(\theta,T)},\label{univrel}
\end{equation}
where $k_{\rm B}$ is the Boltzmann constant and $\tilde{\gamma}(\theta,T)$ defined by
\begin{equation}
\tilde{\gamma}(\theta,T)=\gamma(\theta,T)+\partial^2 \gamma(\theta,T)/\partial \theta^2                                                                         \end{equation}
is the step stiffness.  How the long-range (inverse-square) step-step interaction modifies the relation (\ref{univrel}) is also known.\cite{saam,inverse-square}
Since the essential mechanism of the GMPT form (\ref{GMPTform}) is the non-penetrability of steps, the universality of this form and also of the relation (\ref{univrel}) are fairly ``robust''.  However, for {\em adsorbed} vicinal surfaces,  there may occur breakdown of the GMPT form at a temperature $T_{\rm s}$ where $B(\theta,T)$ vanishes, which we shall show in the present Letter.  Since smallness of $B$ is characteristic of low-temperature stiff surface,  the anomalous behavior at $T_{\rm s}$ may be called {\em stiffening transition}. 

It has been known that adsorbed atoms often change properties of the surface.~\cite{pimpinelli,eaglesham,desjonqueres,copel,jones,fujita,williams98,hannon,latyshev,ozcomert,vonhoegen98} 
\begin{wrapfigure}{r}{6.6cm}
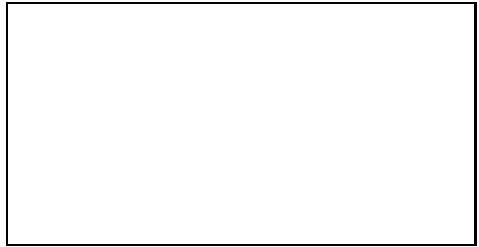

\figurebox{60mm}{3cm}
\caption{ (a) RSOS heights ($h_{1},\ldots,h_{4}$) and edge variables in the mapped vertex model. (b) Decorated-vertex-model representation of the RSOS-Ising coupled system.  Ising spins are represented by circles.}
\label{fig1}
\end{wrapfigure}
  Recently, the adsorption effect on the behavior of step has been studied experimentally.\cite{fujita,williams98,hannon,latyshev,ozcomert,vonhoegen98}  In this Letter, to discuss the adsorption effect on the vicinal surface, we take the restricted solid-on-solid (RSOS) model\cite{RSOS} on the square lattice, coupled with the Ising spin system representing the adsorbed gas.  In the RSOS model, we restrict each nearest-neighbor (nn) height difference $\Delta h$ to be $\Delta h=0,\pm1$, which is a reasonable simplification because configurations with large $|\Delta h|$ are energetically unfavorable.  

We assume that the gas atoms are likely to adsorb at step edge positions\cite{ladeveze} and that the adsorbed atom modifies the ledge energy locally.  The Ising spins are, then, located on the {\em bonds} of the square lattice where the RSOS model is defined;  the Ising spins form a 45$^{\circ}$-rotated square lattice.  We further assume ferromagnetic interactions (attractive interaction, in the lattice-gas picture) with coupling constant $J$ between the nearest-neighbor spins on the rotated square lattice.  We assume simple linear modification of the ledge energy as $\epsilon\rightarrow\epsilon (1-\alpha \sigma)$ ($\sigma$: Ising spin).  This modification leads to the interaction between the RSOS model and the Ising model.  The Hamiltonian of the RSOS-Ising coupled model is, therefore, written as
\beq
{\cal H}=\sum_{<i,j>} \epsilon (1-\alpha \sigma_{b(i,j)}) |h_i-h_j| \nonumber \\
 - J \sum_{<b,b'>} \sigma_{b} \sigma_{b'},
\label{hamil}
\eeq
where $h_i$ is the integer surface height at site $i$, $\epsilon$ the ``bare'' ledge energy, $\sigma_{b(i,j)} = \pm 1$ the Ising spin variable on the bond $b(i,j)$ connecting the nn site pair $<i,j>$.  We should note that the RSOS condition ($|\Delta h| \leq 1$ for each nn site pair) is implicit in (\ref{hamil}). 

We analyze the model by the transfer-matrix method.  For this purpose, we extend the well-known mapping between the RSOS model and the vertex model on the dual lattice,\cite{RSOS,vanBeij} to obtain a ``decorated'' vertex model (Fig.1).  The decorated vertex model can again be regarded as a 6-state vertex model with $19\times16=304$ non-zero vertex weights (304-vertex model).  For approximate diagonalization of the transfer matrix, we employ the product-wavefunction renormalization group (PWFRG) method,\cite{PWFRG,HOA} which is a variant of the White's density matrix renormalization group (DMRG) method\cite{white} (``infinite-system'' algorithm, to be precise).  The PWFRG is specially designed to obtain the fixed point ($=$ thermodynamic limit of the system\cite{Ost-Rom}) of the DMRG efficiently.  For the vicinal surface problem, we have verified the reliability of the PWFRG method\cite{akutsu98}; even with a small number of ``retained bases'' (in the DMRG/PWFRG terminology), the method gives close-to-exact results.

For calculation of $\gamma(\theta,T)$ and $B(\theta,T)$ in (\ref{GMPTform}), we  take a similar approach as we took in Ref. \citen{akutsu98} where we obtained $\gamma$ and $\tilde{\gamma}$ for $2\times1$-reconstructed Si(001).\cite{swartzentruber,bartelt94}  
\begin{wrapfigure}{l}{6.6cm}
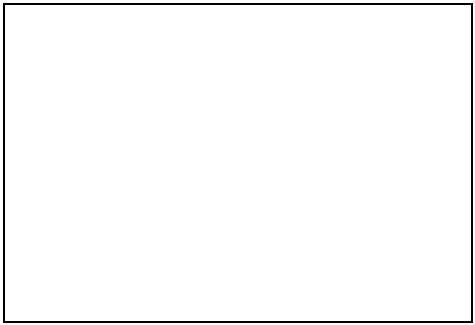

\figurebox{60mm}{4cm}
\caption{
PWFRG results for the $p-\beta \eta$ curves ($\alpha=0.5$ and $J=0.15$).  Temperature of each curve is $\kBT/\epsilon=$ 0.3, 0.35, 0.4, 0.45, and 0.5, from right to left. } \label{fig2}
\end{wrapfigure}
We introduce Andreev field $\eta$\cite{andreev} along the $x$-direction to tilt the surface,  by adding a term $-\eta\sum_{m,n}(h_{(m+1,n)}-h_{(m,n)})$ in the Hamiltonian (\ref{hamil}) ($(m,n)$ is the position vector of the lattice site).  In the vertex-model representation, the surface gradient $p$ along the $x$-direction is just the thermal average of the vertical edge variable of the vertex model, which can easily be calculated from the fixed-point wavefunction obtained by the PWFRG.  By sweeping the field $\eta$, we obtain a $p-\eta$ curve (actual calculation is very similar to that of the magnetization curve for spin chains.\cite{HOA,OHA} See also, Ref. \citen{honda-horiguchi}).  
From the standard argument\cite{andreev} using the Andreev's Legendre transformation $p \rightarrow \eta$, $f(p)\rightarrow \tilde{f}(\eta)=f(p)-p\eta$ ($\eta=\partial f(p)/\partial p$), we have from (\ref{GMPTform}) (with $\rho=p$)
\begin{equation}
\eta = \gamma +3B p^2+ (\mbox{higher order}). \label{eta-p}
\end{equation}
Hence, from the PWFRG calculation, we draw the $\eta-p$ curve and perform the least-square fitting to obtain $\gamma$ and $B$.  For actual fitting, we adopt the fitting form
\begin{equation}
\eta=A_{0}+A_{2}p^2+A_{3}p^3+A_{4}p^4, \label{fit2}
\end{equation}
since the coefficients $A_{3}$ and $A_{4}$ may not be small.  If the relation (\ref{univrel}) holds, we obtain the step stiffness $\tilde{\gamma}(0)$ as
\begin{equation}
\tilde{\gamma}(0)=\frac{\pi^2(k_{\rm B}T)^2}{2A_{2}}.  \label{fit-stiff}
\end{equation}

\begin{wrapfigure}{r}{6.6cm}
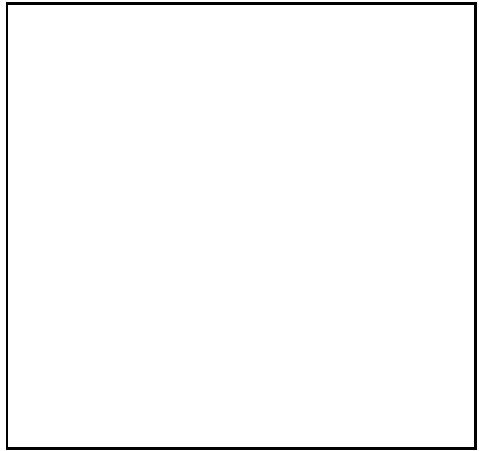

\figurebox{60mm}{5.7cm}
\caption{
Temperature dependence of step stiffness $\tilde{\gamma}(0)$ (triangles) and step tension $\gamma(0)$ (circles) (semi-log plot).  Solid and broken lines are guide to the eyes.
}
\label{fig3}
\end{wrapfigure}
In Fig.2, we show $p$-$\beta \eta$ curves ($\beta=1/(\kBT)$) for $\alpha=0.5$ and $J=0.15$ calculated  by the PWFRG, at some temperatures. 
 In Fig.3, we show temperature dependence of $\tilde{\gamma}(0)$ obtained from (\ref{fit-stiff}).  
In contrast to the ``normal'' cases where $\tilde{\gamma}$ is a monotonically decreasing function of temperature, our $\tilde{\gamma}$ clearly exhibits diverging anomaly at $T_{\rm s}$ ($\kB T_{\rm s}/\epsilon\approx0.4$).  The step tension $\gamma(0)$, on the other hand, behaves smoothly near $T_{\rm s}$, only with weak reentrance.  At $T_{\rm s}$, the $\eta-p$ curve is no longer fitted by the form (\ref{fit2}); in stead, it is well fitted by
\begin{equation}
\eta=A_{0}+A_{4}p^4+A_{5}p^5+\cdots, \label{fit4}
\end{equation}
which is equivalent to a non-GMPT form of expansion
\begin{equation}
f(p)=f(0)+\gamma p+C_5 p^5+(\mbox{higher order}).\label{non-GMPT}
\label{fp5}
\end{equation}
We thus have found vanishing of $B$ at a temperature $T_{\rm s}$ in the RSOS-Ising coupled system, which we call stiffening transition.  As for the $p-\eta$ curve, we should have the change in the critical behavior, from the ordinary square-root type\cite{square-root1} $p\sim(\eta-\eta_{c})^{1/2}$ ($\eta>\eta_c$, $\eta_{c}=\gamma$) to a new one $p\sim(\eta-\eta_{c})^{1/4}$.  We should stress that the stiffening transition has nothing to do with the roughening transition, which takes place well above $T_{\rm s}$ (we estimate $\kB T_{\rm R}/\epsilon \sim 1.35$ in the present case).

We propose two possible mechanisms of the stiffening transition: (1) ``single-particle'' mechanism, and (2) intrinsically many-body mechanism.  The first one is based on  the ``universal free-fermion picture'' of the vicinal surface,\cite{aay88} in a direct way.  In the transfer-matrix treatment for the coarse-grained vicinal surface, the statistical-mechanical problem of many-step system is equivalent to finding the ground state of a free-fermion system in one dimension.  In this view, the ordinary parabolic dispersion of the single-particle energy $\omega(k)\sim \sigma k^2$ (near $k\sim0$) is the source of the $\rho^3$-term in (\ref{GMPTform}) with $B=\sigma\pi^2/3$. Hence, the vanishing of $B$ and appearance of the non-GMPT form (\ref{non-GMPT}) can be simply interpreted as the change of $\omega(k)$ from $k^2$-type to $k^4$-type.  Since $\tilde{\gamma}$ is inversely proportional to the (scaled) squared fluctuation width characterizing the step roughness,\cite{sig-gam} we should have step-smoothening behavior at $T_{\rm s}$.

The second mechanism is based on formation of ``bound states'' of steps.  If we ``integrate out'' the Ising-degrees of freedom, we have an effective RSOS model with effective interactions.  If this interaction amounts to {\em attraction}  between the steps, formation of ``bound state'' (or, ``bound steps'', ``step bunch'') may become possible.  Suppose that steps are forming $n$-body bound states.  We regard each bunch of steps as a ``composite particle'' and assume that the free-fermion picture itself is valid for the system of composite particles.  By $\rho_{n}$, $\gamma_{n}$ and $\tilde{\gamma}_{n}$, we denote the density, formation free energy and the stiffness, of the $n$-body step bunch.  Note that we have $p=n \rho_{n}$, and roughly $\gamma_{n}\sim n\gamma_{1}$ and $\tilde{\gamma}_{n}\sim n \tilde{\gamma}_{1}$.\cite{sudoh}  From a simple dimensional analysis, we see that in  the expansion (\ref{GMPTform}) of the vicinal-surface free energy $f(p)$, the coefficient $B$  scales as $B\sim 1/n^4$.  Hence, if $n$ diverges then $B$ vanishes; $T_{\rm s}$ is the condensation temperature of the steps.

To summarize, we have discussed the adsorption effect on the vicinal surface below the roughening temperature, in terms of the restricted solid-on-solid model coupled with the Ising model.  By employing the product-wavefunction renormalization group method, we have obtained temperature dependence of the step tension and step stiffness.  We have found a stiffening transition where the step interaction coefficient vanishes, and at the same time, the vicinal surface free energy takes a different form from the Gruber-Mullins-Pokrovsky-Talapov-type form.

We should note that anomalous increase of step stiffness has been observed\cite{hannon} for boron (B)-doped Si(001).  Similar phenomenon has also been observed\cite{latyshev} for high temperature Si(111) where Si adatom layer is known to exist.\cite{iwasaki,khomoto,latyshev91,yang94}  The stiffening transition we have found may have relevance to these anomalous behavior.  In this Letter, we have restricted the analysis to special (but typical, in our view) set of parameter values of the model.  To explore full parameter space is an important problem, which may lead to discovery of other interesting phenomena and may also lead to clarification of the mechanism of the stiffening transition.  Models with other lattice structures, and/or with other types of couplings are also to be explored, for quantitative explanation of experiments.

The authors thank  Prof. A. V. Latyshev, Prof. T. Koshikawa, Prof. T. Yasue, Prof. A. Ichimiya, and Dr. N. C. Bartelt for  helpful discussions.  Thanks are also due to  Prof. T. Nishinaga and Prof. H. Iwasaki for encouragement.  This work was partially supported by  the ``Research for the Future'' Program  from The Japan Society for the Promotion of Science (JSPS-RFTF97P00201) and by the Grant-in-Aid for Scientific Research from Ministry of Education, Science, Sports and Culture (No.09640462).

\end{document}